\documentstyle[epsf,12pt]{article}

\begin{document}

\noindent

\begin{center}

{\Large\sf Search for the radiative capture $d+d \rightarrow
^4\!He+\gamma $ reaction from the $dd\mu$ muonic molecule state.}

\vspace{1cm}

L.N.~Bogdanova$^1$, V.R.~Bom$^2$,
 D.L.~Demin,
C.W.E.~van~Eijk$^2$, V.V.~Filchenkov, N.N.~Grafov, V.G.~Grebinnik,
K.I.~Gritsaj, A.D.~Konin, A.V.~Kuryakin$^3$, V.A.~Nazarov$^3$,
V.V.~Perevozchikov$^3$, A.I.~Rudenko, S.M.~Sadetsky$^4$,
Yu.I.~Vinogradov$^3$ A.A.~Yukhimchuk$^3$, S.A.~Yukhimchuk,
V.G.~Zinov, S.V.~Zlatoustovskii$^3$

\end{center}

\vspace{1cm}

{Joint Institute for Nuclear Research (JINR), Dzhelepov Laboratory
of Nuclear Problems, Dubna, 141980  Russia}

{$1-$ State Scientific Center of Russian Federation,\\ Institute
of Theoretical and Experimental Physics, Moscow, 117218}

{$2-$ Delft University of Technology, 2629 JB Delft, the
Netherlands}

{$3-$ Russian Federal Nuclear Center, All-Russian Research
Institute
      of Experimental Physics (RFNC-VNIIEF),
Sarov, Nizhny Novgorod reg., 607200}

{$4-$ St.~Petersburg Nuclear Physics Institute (PNPI), Gatchina,
     188350}

\vspace{1cm}

\begin{abstract}
A search for the muon catalyzed fusion reaction $dd \rightarrow
 ^4\!He +\gamma$ in the $dd\mu$ muonic molecule was performed
 using the experimental $\mu CF$ installation TRITON and
$NaI(Tl)$ detectors for $\gamma$-quanta. The high pressure target
filled with deuterium at temperatures  from 85 K to 800 K was
exposed to the negative muon beam of the JINR phasotron to detect
$\gamma$-quanta with energy 23.8 MeV.
 The first experimental estimation for the yield of the radiative
deuteron capture from the $dd\mu$ state J=1 was obtained at the
level $n_{\gamma}\leq 2\times 10^{-5}$ per one fusion.
\end{abstract}

\section{Introduction}

It is understood that investigations of the fusion reactions
between hydrogen isotope nuclei at low energies are of great
importance for determining properties of lightest nuclei and for
the astrophysics. In particular, there is a need for new or
improved measurements of many radiation capture reactions included
in various astrophysical scenarios.
 Due to the Coulomb repulsion fusion cross-sections $\sigma(E)$
drop rapidly at low ($E\leq 100 keV$) collision energies (in an
exponential scale for "bare" nuclei).

Reaction
\begin{equation}
d+d \longrightarrow ^4\!He+\gamma +23.8\, MeV
\end{equation}
 is involved in both primordial and
stellar nucleosynthesis.  Its cross section is rather small (about
1 pb at 50 keV, compared to 1 mb for main fusion channels
$d(d,n)^3He$ and $d(d,p)^3H$), and its experimental investigations
in $dd$ collisions are rather difficult.

At energies $E > 400$ keV reaction (1) proceeds  mainly by a
d-wave E2 transition to the $^1S_0$-state of $^4$He \cite{Wilk}.
 The reason is the identical boson character of the entrance channel
requiring $L +S$ to be even ($L$ and $S$ are orbital angular
momentum and total spin of the $dd$ system).   At lower energies,
the centrifugal barrier suppresses the d-wave E2 capture, allowing
s-wave E2 transition to the D-state admixture of $^4$He.
 Measurements extended to energies
below 100 keV \cite{Barn} have confirmed  this picture. However,
an  existence of multipoles other than E2 in reaction (1) was not
excluded experimentally despite belief that dipole transitions
$E1$ and $M1$ with $\Delta S=0$ should be suppressed due to
standard isospin selection rule $\Delta T=0$.

 Measurements of the cross section angular distributions
 $\sigma(\theta)$, of vector $A_y$
and tensor $A_{yy}$ analyzing powers performed with a polarized
deuteron beam with energy $E_d$(lab) = 80 keV  stopping in the
target  have yielded an unexpected observation of the p-wave
strength in $^2H(\vec{d},\gamma)^4He$ reaction \cite{Kram}. It has
been found that over 50\% of the cross section strength at these
low energies is due to E1 and M2 p-wave capture. This finding
might affect the low energy behavior of the S-function and be
considered as an isospin selection rule violation.(Some other
evidences for non-E2 radiation are cited in \cite{Kram}). It would
be extremely interesting to observe this p-wave manifestation in
an independent measurement.

During past decades experiments in which various fusion reactions
between hydrogen isotopes are catalyzed by muons have provided
supplementary information about these reactions at energies well
below the lowest energies accessible by conventional beam-target
experiments \cite {Bog}. In the muon catalyzed (MC) process fusion
takes place from the bound states of muonic molecules.  Nuclei inside
muonic molecules are practically at rest, being
 separated by average distances
$a_{\mu}\sim {\hbar}^2/e^2 m_{\mu}^2=2.5\cdot10^{-11}\,cm$ (
$m_{\mu}$ is the muon mass).

 Muonic molecules can be formed
in the states with total angular momenta $J=0$ and $J=1$, that
correspond to the relative orbital angular momenta of nuclei $L=0$
and $L=1$. Depending on the hydrogen isotope mixture parameters
various states of muonic molecules can be populated. This makes
possible study fusion reactions at super-low energies from
prepared  $s-$ and $p-$ nuclear states with definite spins.

Study of the MC fusion process in $dd\mu$ muonic molecule resulted
in the complimentary and detailed information about
charge-symmetric reactions $d(d,n)^3He$ and $d(d,p)^3H$. A
significant difference in the p-wave parts of the $d(d,p)^3H$ and
$d(d,n)^3He$ reaction yields was observed in the experiments with
low energy polarized deuteron beams \cite {Ad}. Comparison of two
reaction branches showed some s-wave enhancement together with
essential p-wave suppression of the proton branch. (This result
was then interpreted by some as an evidence for charge symmetry
breaking forces.)

Direct measurement of the yields ratio $R_p(n/p)$ for reactions
\begin{equation}
dd\mu \longrightarrow ^3\! He+n+\mu, \, \mu ^3\!He + n,
\end{equation}
\begin{equation}
dd\mu \longrightarrow t+p +\mu
\end{equation}
proceeding from the $J=1$ state of $dd\mu$ molecules \cite{Ba},
\cite{Gatch} gave the value $R_p(n/p) = 1.42 \pm 0.03 $. It agreed
with the ratio from \cite {Ad} determined from the elaborate (and
model dependent) analysis of the in-flight data. Rates of $dd\mu$
fusion reactions (2),(3) from the p-wave were experimentally
measured \cite {lfdd} and corresponding nuclear reaction constants
were extracted from MC data \cite{Hale}.

The deuteron radiative capture reaction in the $dd\mu$-molecule
\begin{equation}
dd\mu \longrightarrow ^4\!He \mu+\gamma +23.8\,MeV,
\end{equation}
has not previously been investigated because of the extreme
smallness of its expected yield. In the systematic study of the MC
process in deuterium we have recently performed \cite{OurlastDD}
measurements in the temperature range T=85-800 K.  As in our
earlier experiments \cite {PrDD}, neutrons from reaction (2) were
detected. At temperatures $T > $ 150 K $dd\mu$ molecules are
mainly formed in J=1 state and fusion reactions proceed from the
p-wave of relative nuclear motion. Hence, being detected, 23.8 MeV
$\gamma$-quanta would unambiguously indicate a finite p-wave
contribution to process (4) rate.

 In view of this, we performed a
feasibility study of process (4) detection in our
last measurements of the $dd\mu$-molecule formation rate
\cite{OurlastDD}. For
this aim one of two usually used neutron detectors \cite{PrDD}was
 removed and a gamma detector was installed instead. The
level of the radiation  background in our installation has been
measured. We present the first experimental estimation for the
yield of the radiative deuteron capture in the p-wave from the
$dd\mu$-molecule.

\section{Experimental setup}

 The experimental setup (its layout is shown in Fig.~1) has been
described in detail in \cite{Novel}. The muon beam of the JINR
phasotron was directed into a high pressure deuterium target
(HPDT) \cite{Targ}.  Scintillation counters 1-4 detected incoming
muons. Cylinder-shaped counter 5 served to identify the muon stop
in the target and to detect electrons from muon decay. A
coincidence between signals of counters $5$ and $1e,2e$ served as
the electron signal.

A large neutron detector ND (volume of NE-213 $v=12.5\,l$)
\cite{Nsp}, \cite{Novel} was aimed to detect neutrons from
reaction (2). To reduce a background, the $n-\gamma$ separation
was realized by comparing the signals for the total charge and the
fast component charge of the ND pulse. The efficiency of the
$\gamma$-quanta discrimination was better than $10^{-3}$ for
energies $E_{\gamma, e}>100\,keV$.

\bigskip
\begin{center}
\epsfclipon
\mbox{\epsfysize=60mm\epsffile{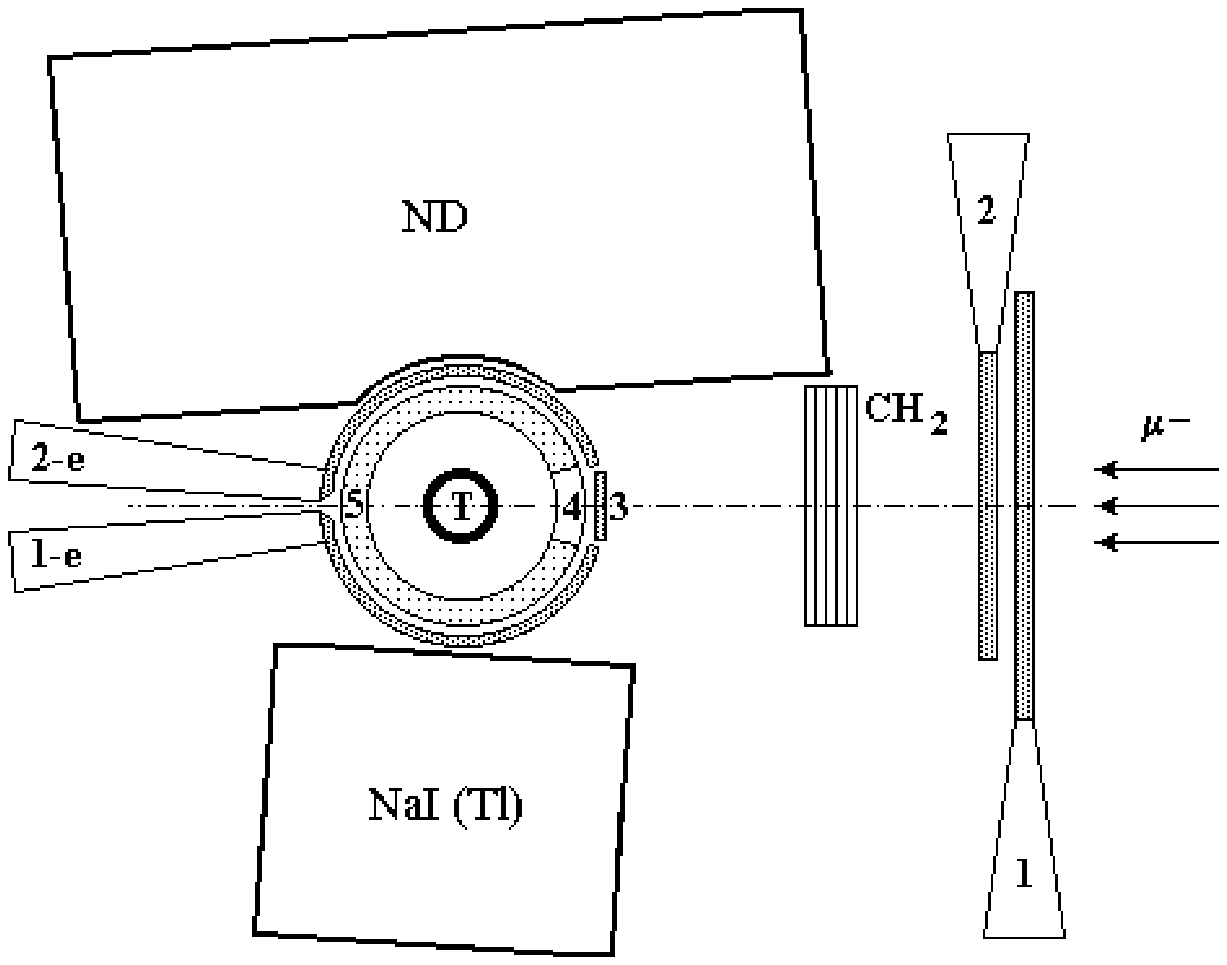}}
\hspace{1cm}
\begin{minipage}[b]{60mm}
Figure 1. Experimental layout.
\\[20mm]
\end{minipage}
\end{center}
 The $\gamma$-quanta were detected with a
$NaI(Tl)$ crystal of a 150 mm diameter and a 100 mm height.
 It was calibrated with $\gamma$-sources
$^{60}Co$ (total energy of two $\gamma$-s  2.5 MeV), $Pu$-$Be$
($E_{\gamma}=4.43$\,MeV) and with 5.5 MeV $\gamma$-s from the
reaction
\begin{equation}
pd\mu \rightarrow ^3\! He\mu+\gamma.
\end{equation}
 Reaction (5) was observed in the test exposures when
the target was filled with H/D mixture containing about 20\% of
protium. The energy resolution of the detector was $15\%$ FWHM at
5.5 MeV.

Linearity of the  energy scale was checked under different
voltages supplied to the detector in the measurements with
available $\gamma$-sources $^{60} Co$ and $Pu$-$Be$. In the used
amplitude region it proved to be linear at the level  2-3\%.
Expected position for $\gamma$-s from (4) is then approximately
200-th channel.  Stability of the spectrometric system was
controlled with $\gamma$-sources during the run.

We estimated the $\gamma$-quanta detection efficiency using values
of $\gamma$-s cross sections in $NaI$ and the known solid angle of
the detector. Taking into account efficiency losses ($30-40\,\%$)
due to the bremsstrahlung in the target walls we determined the
efficiency for 24 MeV $\gamma$-quanta detection
\begin{equation}
\epsilon _{\gamma} = (5 \pm 1)\%.
\end{equation}

 Primary selection of the events detected by the neutron and
 $\gamma$-detectors was realized by the trigger. It allowed
 only those events for further time and amplitude analysis
which were connected with electron registration, i.e., delayed
coincidences $\mu - n,\gamma - e$ were used. Under this condition
the timing sequence of the $NaI$ and ND signals was registered by
flashes ADC and recorded on the PC.

\section{Measurements}
During the run eight exposures were performed at different
deuterium temperatures and densities. Experimental conditions for
them are presented in Table 1. Deuterium density is given in
relative units: $\phi=n/n_0$ where $n_0 =4.25 \cdot
10^{22}\,nucl/cm^{3}$ is the liquid hydrogen density (LHD).
For all exposures the intensity of muons detected by counters 1-4
was $2.5 \cdot 10^3\,s^{-1}$. The electron counting rate was 15-30
per second depending on the deuterium density.

\begin{center}

Table 1. Experimental conditions.

\bigskip

\begin{tabular}{|c|c|c|c|c|c|c|c|}

\hline

Exposure & T, K &\multicolumn{2}{c|}{Content, \%}& $\phi,$ & $N_e$
& $N_n$&$N_{dd\mu}$ \\

\cline{3-4}

         &      &$C_p$(H)            &$C_d$(D)                & $LHD$  &       &  &     \\

\hline

1&  85 (5) &20.7 (0.5)&79.3 (0.5)& 0.84 (0.03) & 712~300 & 4~000& $1.2 \cdot 10^5$\\

\hline

2& 110 (5) &20.7 (0.5)&79.3 (0.5)& 0.84 (0.03) & 474~600 & 4~700& $1.2 \cdot 10^5$\\

\hline

3& 230 (5) &20.7 (0.5)&79.3 (0.5)& 0.83 (0.03)& 433~200 & 15~000& $4.5 \cdot 10^5$\\

\hline

4& 301 (4) &20.7 (0.5)&79.3 (0.5)& 0.83 (0.03)& 443~700 & 20~200& $6.1 \cdot 10^5$\\

\hline

5& 299 (4) &20.7 (0.5)&79.3 (0.5)& 0.47 (0.02)& 388~900 & 13~900& $4.2 \cdot 10^5$\\

\hline

6& 298 (4) & 0.1 (0.1)&99.9 (0.1)& 0.50 (0.02)& 232~500 & 18~100& $5.7 \cdot 10^5$\\

\hline

7& 548 (10)& 0.1 (0.1)&99.9 (0.1)& 0.50 (0.02)& 240~000 & 19~500& $5.1 \cdot 10^5$\\

\hline

8& 791 (15)& 0.1 (0.1)&99.9 (0.1)& 0.49 (0.02)& 315~000 & 20~500& $6.1 \cdot 10^5$\\

\hline

\end{tabular}

\end{center}

\vspace{1cm}

 The number of detected electrons
from muon decay, $N_e$,  was determined from the fit of the electron
time spectra taking into account the background from muon stops in
the target walls. The latter one was found in the exposure with
the empty target. Number of electrons
detected per 10 hours of phasotron operation (one exposure) was
$\simeq (0.5-1.0) \cdot 10^6$.
Details of the  analysis can be found in
\cite{PrDD}.

The number of neutrons from reaction (2), $N_n$, was obtained from
the analysis of the time spectra of the events detected by ND and
belonging to the neutron region in the $n-\gamma$ plot \cite{Nsp}.
The neutron background was measured in the special exposure with
the target filled with helium. Thus determined numbers of $N_e$
and $N_n$ are presented in Table 1.

Exposures made with H/D mixture allowed detection of reaction (5),
which was used both  for energy calibrating and checking the
$\gamma$-quanta detection efficiency.

As expected, exposures 1 and 2 are characterized by a low
neutron/electron ratio. In other words, only small fraction of the
muon stops in the target lead to formation of $dd\mu$-molecules
and subsequent reaction (2), detected in our experiment.
It is due to the fact
\cite{Ba}, \cite{Gatch}, \cite{lfdd}, \cite{PrDD} that,
at low temperatures $T < 150 \,K$,
the $dd\mu$ formation rate
${\lambda}_{dd\mu} \cdot \phi <0.1\cdot 10^6\,s^{-1}$ is
small  compared to
the $d\mu$-atom disappearance rate
${\lambda}_{d\mu}={\lambda}_0 + {\lambda}_{dd\mu}\cdot \phi
\cdot(1-C_p)+
{\lambda}_{pd\mu} \cdot  \phi \cdot C_p$, where
 ${\lambda}_0=4.55 \cdot 10^5\,s^{-1}$ is the free muon
disappearance rate and ${\lambda}_{pd\mu} =(5.53 \pm 0.16) \cdot
10^6 \,s^{-1}$ \cite{PrPd} is the $pd\mu$ formation rate. This
allowed use these exposures for the estimate of the accidental
background. Exposures 3-8 were accepted for the search of
$\gamma$-s from reaction (4) and estimation of its relative yield.

From the measured numbers of neutrons $N_n$, known efficiency of
neutron detection ${\epsilon}_n=13\,\%$ \cite{Bom} and partial
probability of reaction (2) $\beta \cong 0.58$ \cite{Ba}
\cite{Gatch}
 $dd\mu$ formation rates
${\lambda}_{dd\mu}$ were determined for each exposure. With thus
obtained values $\lambda _{dd\mu}$ and data from Table~1 we
calculated average number of catalysis cycles, $n_c$, using a
simplified formula describing $\mu$CF kinetics in H/D mixture
(see, e.g. \cite{Peti})
\begin{equation}
n_c={\lambda}_{dd\mu} \cdot \phi \cdot (1-C_p)/
[{\lambda}_0+{\lambda}_{dd\mu} \cdot \phi \cdot (1-C_p)\cdot
{\varpi}_{dd} +{\lambda}_{pd\mu} \cdot \phi \cdot C_p \cdot
{\varpi}_{pd}].
\end{equation}
Here ${\varpi}_{dd}=0.07$ and ${\varpi}_{pd}=0.85$ are branching
ratios of fusion reactions with muon sticking to helium with
respect to all fusion channels in muonic molecules $dd\mu$ and
$pd\mu$. These reactions lead to the muon loss from the catalysis
cycle. Using thus determined $n_c$ and measured numbers of
electrons $N_e$ we could calculate numbers of $dd\mu$ molecules
$N_{dd\mu}$, formed in each exposure, as $N_{dd\mu}=N_e \cdot
n_c$. The results are presented in Table~1 and the total number of
$dd\mu$ molecules for exposures 3-8  was used for the estimation
of reaction (4) yield.

\section{Analysis of $\gamma$-events}
Of all events detected by the $\gamma$-detector those with
$\gamma$ energy
\begin{equation}
E_{\gamma}>17\,MeV
\end{equation}
were selected for the further analysis.   These events accumulated
in exposures 3-8 were displayed in a two-dimensional plot $\gamma$
time ($t_{\gamma}$) - electron time ($t_e$), shown in Fig.~2.

\vspace{1cm} \hspace{-1cm}
\mbox{\epsfysize=80mm\epsffile{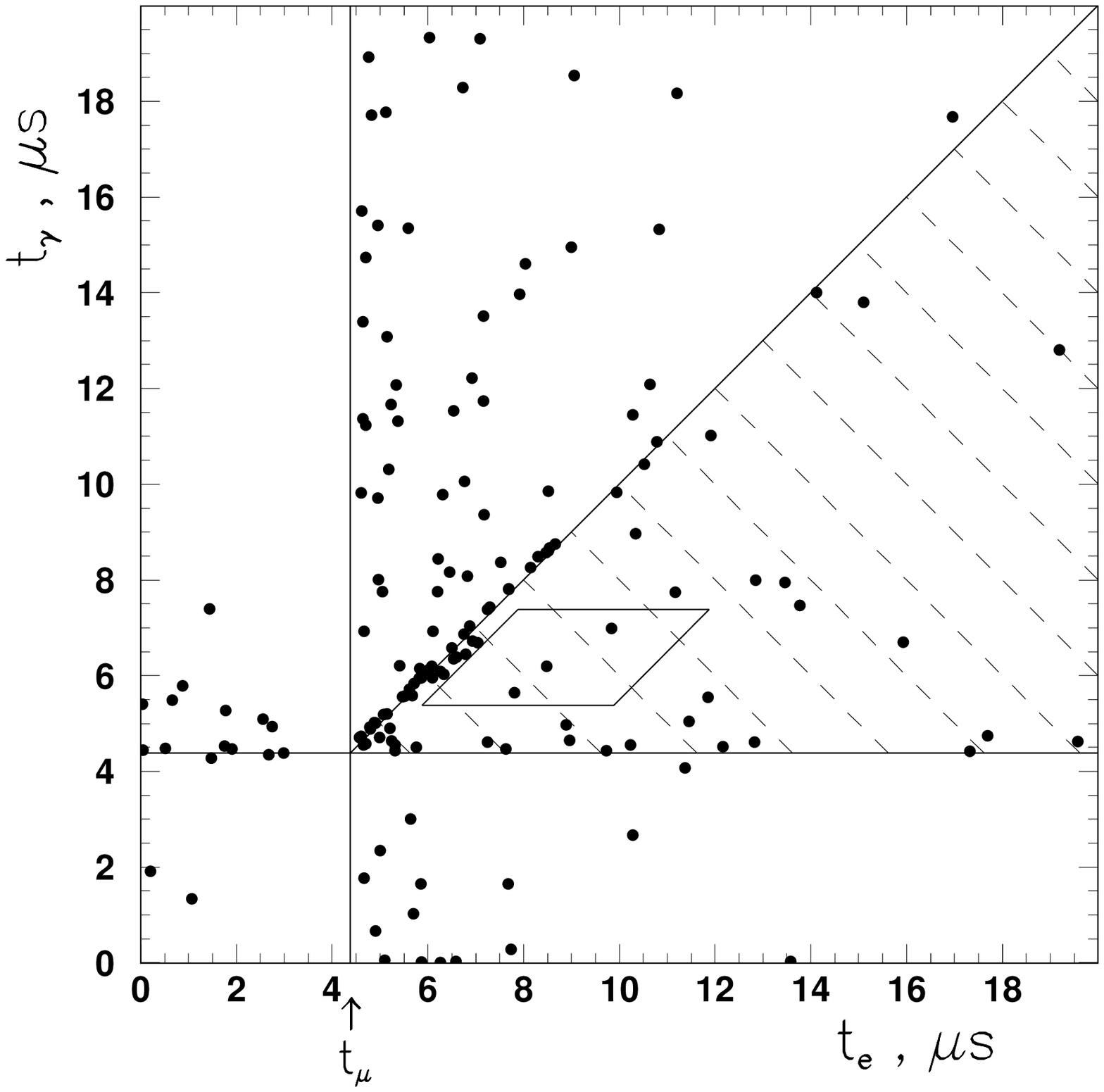}} \hspace{1cm}
\epsfclipon
\begin{minipage}[b]{60mm}

Figure 2. Two-dimensional $t_e -t_{\gamma}$ plot for the events
with $E_{\gamma} >17$ MeV detected by $NaI$ detector in exposures
3-8.
\\[10mm]
\end{minipage}

\bigskip

Fusion events from reaction (4) should arrive after the muon
entrance ($t_{\mu}$) and before  the muon decay ($t_e$), so for
the  primary selection the following time sequence
$t_e,\,t_{\gamma}>t_{\mu}$, $t_e >t_{\gamma}$ was required,
corresponding to the  dashed area in Fig.~2.

It is seen that a noticeable fraction of events in this plot is
concentrated at small $t_e-t_{\mu}$, $t_{\gamma}-t_{\mu}$. These
events might be a manifestation of the muon stops in the target
walls. In their material (Ni, Fe) muon disappears after the
average time ${\tau}_{\mu}=0.2\,\mu s$, either due to its decay
starting the false trigger, or due to the nuclear capture with
emission of capture products. To reduce the background originating
from such processes events corresponding to fast $\gamma$- and
electron emission should be excluded from the consideration by
introducing a time delay with respect to $t_{\mu}$.

From the other side, time distribution of events resulting from
the $dd\mu$ molecule fusion (4) should obey the exponential law
 $$f_{\gamma}(t)=Const \cdot exp([-{\lambda}_0+\phi
{\lambda} _{dd\mu} {\varpi}_{dd}]t),$$ so allowing a large time
delay would lead to the loss of the efficiency.

To  provide the suppression of the accidental background and
simultaneously avoid the efficiency losses the following time
intervals were chosen:
\begin{equation}
1\,\mu s < t_{\gamma} - t_{\mu} < 3\,\mu s ; \,\,\,
 0.5 \mu s < t_{\gamma} - t_e  <  4.5 \, \mu s.
\end{equation}
The corresponding  region is indicated by a BOX  in Fig.~2 with 3
$N_{\gamma}^t$ events inside.

To estimate the background, we selected the  area
 $t_{\gamma}>1\,\mu s$,
$0.5\,\mu s < t_e < 4.5\,\mu s$ and found 7 events there. This
corresponds to the number of background events in the region (9)
$N_{\gamma}^{ b1}=2\pm 1$.

In addition, for an independent estimation of the accidental
background, events from exposures 1,2 satisfying to (8) and (9)
were selected. The number of such events normalized to the number
of electrons in exposures 3-8 was found $N_{\gamma}^{ b2}=2$. It
proved to be at the level of the previous estimate obtained from
exposures 3-8.

The background is found to exceed the measured intrinsic
background of the installation, corresponding to the cosmic ray
intensity at sea level, by a factor of 2. We conclude that
additional background is  correlated with phasotron operation.

 The energy distribution of the events detected by the $NaI$
detector and selected with criteria (9) for exposures 3-8 (solid
line) is shown in Fig.~3. The dashed line is the spectrum for the
normalized background.

\bigskip

\epsfclipon \mbox{\epsfysize=60mm\epsffile{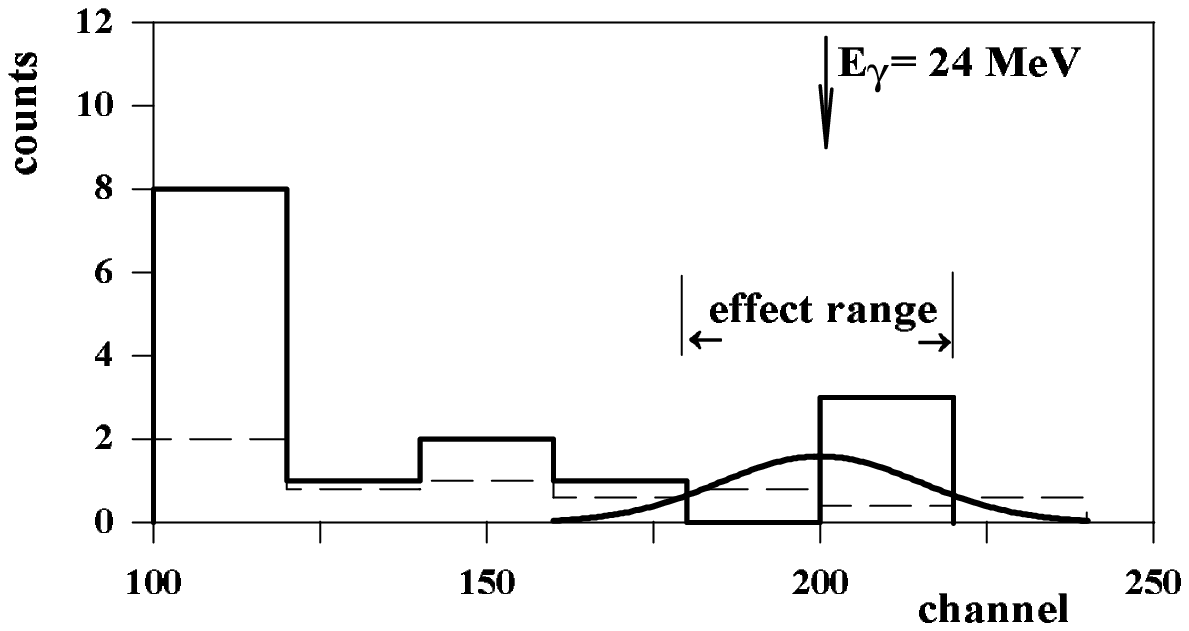}}

Figure~3. Amplitude $\gamma$-quanta spectra for the events
selected with criteria (8) for exposures 3-8 (solid line) and the
normalized "background" ones (dashed line). Response function of
the $NaI$ detector is represented by the gaussian.

\bigskip

It is seen from the figure that these spectra  practically
coincide for energies $E_{\gamma}>17$ MeV. Some excess of events
for lower energies can be ascribed to the background induced by
neutrons from reaction (2).

 From considerations above the number
of candidate events can be obtained
\begin{equation}
N_{\gamma}=N_{\gamma}^t-N_{\gamma}^b=1\pm 2
\end{equation}

The measured yield of reaction (4) per $dd\mu$-molecule is
evaluated as
\begin{equation}
{\eta}_{\gamma}= \frac{ N_{\gamma}}{{\epsilon}_{\gamma} \cdot
N_{dd\mu}^{tot}},
\end{equation}
where $N_{dd\mu}^{tot}$ is the total number of $N_{dd\mu}$
molecules accumulated in exposures 3-8:
\begin{equation}
N_{dd\mu}^{tot}=3.4 \cdot 10^6
\end{equation}

Using the estimation (6) for the efficiency of $\gamma$-quanta
registration and taking into account the selection efficiency due
to the accepted criteria (9) one obtains the detection efficiency
of $\gamma$-s from  reaction (4)
\begin{equation}
{\epsilon}_{\gamma} = (3 \pm 0.5)\, \%
\end{equation}

Substituting values (10), (13) and (14) into Eq.~(11) we obtain
for the absolute $\gamma$ yield per one $dd\mu$-molecule
\begin{equation}
{\eta}^{(1)} _{\gamma}=(1 \pm 2) \cdot \,10^{-5}.
\end{equation}

In the second run new measurements with deuterium target and
$NaI(Tl)$ detector of larger size have been conducted and result
of the similar analysis of the new data set is in agreement with
(14)
\begin{equation}
{\eta}^{(2)} _{\gamma}=(0.8 \pm 1.5) \cdot 10^{-5}.
\end{equation}

Combining (14) and (15) one obtains (at  90\% C.L.)
\begin{equation}
{\eta}_{\gamma}< 2 \cdot 10^{-5}.
\end{equation}

From here an upper limit for the radiative fusion rate
${\lambda}_{\gamma}^1$ from the J=1 state of $dd\mu$ molecule can
be deduced using the experimental value of the total fusion rate
${\lambda}_f^1=4 \cdot 10^8\, s^{-1}$ \cite{Hale} $$
{\lambda}_{\gamma}^1 <8 \cdot 10^{3}\,s^{-1}.$$

\section{Conclusion}

The first attempt has been undertaken to estimate the yield of
radiative capture  reaction (4) from the J=1 state of $dd\mu$
muonic molecule. The background conditions were evaluated and
appropriate methods of data analysis were elaborated. The
sensitivity of the present experiment is not enough to make a
decisive conclusion about the p-wave contribution to the process
of radiative $dd$ capture. (With the data from \cite{Kram} we
would expect ${\eta}_{\gamma} \sim 10^{-6}$.) We plan to achieve
this level of sensitivity using new ${\gamma}$-detectors of larger
efficiency. Of crucial importance is the understanding of the
background structure and elaboration of background suppression
methods.

\section{Acknowledgements}
This work has been supported by the Department of Atomic Science
and Technology of Minatom of Russia under the Contract ¹
6.25.19.19.00.969. The  authors thank L.I. Ponomarev for support
and  V.B.~Belyaev, V.E.~Markushin and L.N.~Strunov for the
stimulating discussions.

\end{document}